\documentstyle[12pt,setspace,epsf]{article}
\def \Et {{\rm E}_{\rm T}}
\def \Pt {{\rm P}_{\rm T}}

\newcommand{\MET}{\mbox{$\protect \raisebox{.3ex}{$\not$}\et$}}

\def\Z0{${\em Z^0\/}$}

\def\roots{${\sqrt s}\:$}
\def\r#1 {$^{#1}$}

\newcommand{\et}{{\rm E}_{\scriptscriptstyle\rm T}}

\newcommand{\met}{\mbox{$\protect \raisebox{.3ex}{$\not$}\et$}}

\newcommand{\ttbar}{t\bar{t}}
\newcommand{\bbbar}{b\bar{b}}
\newcommand{\ccbar}{c\bar{c}}

\def \Mt {M_{top}}

\newcommand{\ipb}{ {\rm pb}^{-1} }
\newcommand{\degs}{\mbox{$^{\circ}$}}

\def\gepsfcentered#1{
  \def\testit{#1}
  \def\lbracket{[}
  \ifx\testit\lbracket
    \let\dofilecmd=\gepsfwithopt
  \else
    \let\dofilecmd=\gepsfnoopt
  \fi
  \dofilecmd}

\def\gepsfnoopt#1{
  \begin{center}
  \leavevmode
  \epsffile{#1}
  \end{center}}

\def\gepsfwithopt#1 #2 #3 #4]#5{
  \begin{center}
  \leavevmode
  \gepsfmaxx=0.94\textwidth
  \epsffile[#1 #2 #3 #4]{#5}
  \end{center}}

\newdimen\gepsfmaxx
\def\epsfsize#1#2{
  \ifnum \epsfxsize=0
    \ifnum \epsfysize=0
      \ifnum #1 > \gepsfmaxx
        \gepsfmaxx
      \else
        #1
      \fi
    \else
      \epsfxsize
    \fi
  \else
    \epsfxsize
  \fi
}
\begin{document}
\begin{flushright}
\singlespacing
FERMILAB-PUB-95/022-E \\
CDF/PUB/TOP/PUBLIC/3040 \\
\end{flushright}

%

\doublespacing
\begin{center}
{\large Observation of Top Quark Production in $\bar{p} p$ Collisions}
\end{center}

\begin{abstract}
We establish the existence of the top quark using a 67 $\ipb$ data sample
of $\bar{p} p$
collisions at \roots = 1.8 TeV collected with the Collider Detector at
Fermilab (CDF).
Employing techniques similar to those we previously published,
we observe a signal consistent with $\ttbar$ decay to $WWb\bar{b}$,
but inconsistent with
the background prediction by $4.8\sigma$.  Additional evidence for
the top quark is provided by a peak in the reconstructed mass distribution.
We measure the top quark mass to be
$176\pm8({\rm stat.})\pm10({\rm sys.})$~GeV/c$^2$, and the $\ttbar$
production cross section to be $6.8^{+3.6}_{-2.4}$ pb.

\end{abstract}

\vspace{0.5cm}

\begin{center}
	               {\em The CDF Collaboration} \\
\font\eightit=cmti8
\def\r#1{\ignorespaces $^{#1}$}
\hfilneg
\begin{sloppypar}
\noindent
F.~Abe,\r {14} H.~Akimoto,\r {32}
A.~Akopian,\r {27} M.~G.~Albrow,\r 7 S.~R.~Amendolia,\r {24}
D.~Amidei,\r {17} J.~Antos,\r {29} C.~Anway-Wiese,\r 4 S.~Aota,\r {32}
G.~Apollinari,\r {27}
% ...
\end{sloppypar}

\vskip .025in
\begin{center}
(CDF Collaboration)
\end{center}

\vskip .025in
\begin{center}
\r 1  {\eightit Argonne National Laboratory, Argonne, Illinois 60439} \\
\r 2  {\eightit Istituto Nazionale di Fisica Nucleare, University of Bologna,
I-40126 Bologna, Italy} \\
\r 3  {\eightit Brandeis University, Waltham, Massachusetts 02254} \\
\r 4  {\eightit University of California at Los Angeles, Los
Angeles, California  90024} \\
\r 5  {\eightit University of Chicago, Chicago, Illinois 60637} \\
\r 6  {\eightit Duke University, Durham, North Carolina  27708} \\
\r 7  {\eightit Fermi National Accelerator Laboratory, Batavia, Illinois
60510} \\
\r 8  {\eightit Laboratori Nazionali di Frascati, Istituto Nazionale di Fisica
               Nucleare, I-00044 Frascati, Italy} \\
\r 9  {\eightit Harvard University, Cambridge, Massachusetts 02138} \\
\r {10} {\eightit Hiroshima University, Higashi-Hiroshima 724, Japan} \\
\r {11} {\eightit University of Illinois, Urbana, Illinois 61801} \\
\r {12} {\eightit Institute of Particle Physics, McGill University, Montreal
H3A 2T8, and University of Toronto,\\ Toronto M5S 1A7, Canada} \\
\r {13} {\eightit The Johns Hopkins University, Baltimore, Maryland 21218} \\
\r {14} {\eightit National Laboratory for High Energy Physics (KEK), Tsukuba,
Ibaraki 305, Japan} \\
\r {15} {\eightit Lawrence Berkeley Laboratory, Berkeley, California 94720} \\
\r {16} {\eightit Massachusetts Institute of Technology, Cambridge,
Massachusetts  02139} \\
\r {17} {\eightit University of Michigan, Ann Arbor, Michigan 48109} \\
\r {18} {\eightit Michigan State University, East Lansing, Michigan  48824} \\
\r {19} {\eightit University of New Mexico, Albuquerque, New Mexico 87131} \\
\r {20} {\eightit Osaka City University, Osaka 588, Japan} \\
\r {21} {\eightit Universita di Padova, Istituto Nazionale di Fisica
          Nucleare, Sezione di Padova, I-35131 Padova, Italy} \\
\r {22} {\eightit University of Pennsylvania, Philadelphia,
        Pennsylvania 19104} \\
\r {23} {\eightit University of Pittsburgh, Pittsburgh, Pennsylvania 15260} \\
\r {24} {\eightit Istituto Nazionale di Fisica Nucleare, University and Scuola
               Normale Superiore of Pisa, I-56100 Pisa, Italy} \\
\r {25} {\eightit Purdue University, West Lafayette, Indiana 47907} \\
\r {26} {\eightit University of Rochester, Rochester, New York 14627} \\
\r {27} {\eightit Rockefeller University, New York, New York 10021} \\
\r {28} {\eightit Rutgers University, Piscataway, New Jersey 08854} \\
\r {29} {\eightit Academia Sinica, Taipei, Taiwan 11529, Republic of China} \\
\r {30} {\eightit Texas A\&M University, College Station, Texas 77843} \\
\r {31} {\eightit Texas Tech University, Lubbock, Texas 79409} \\
\r {32} {\eightit University of Tsukuba, Tsukuba, Ibaraki 305, Japan} \\
\r {33} {\eightit Tufts University, Medford, Massachusetts 02155} \\
\r {34} {\eightit University of Wisconsin, Madison, Wisconsin 53706} \\
\r {35} {\eightit Yale University, New Haven, Connecticut 06511} \\
\end{center}
\end{center}
\vspace{0.5cm}


%

Recently CDF presented the first direct evidence for the top
quark\cite{prdl},
the weak isodoublet partner of the $b$ quark required in the Standard Model.
We searched for
$\ttbar$ pair production with the subsequent decay
$\ttbar\rightarrow WbW\bar{b}$.
The observed topology in such events is determined by the decay mode of the
two W bosons.  Dilepton events ($e\mu$, $ee$, $\mu\mu$) are produced primarily
when both W bosons decay into $e\nu$ or $\mu\nu$.  Events in the
lepton+jets channel ($e,\mu$+jets) occur when one W boson decays into leptons
and the other decays into quarks.  To suppress background in the lepton+jets
mode, we identify $b$ quarks by reconstructing secondary vertices from
$b$ decay (SVX tag) and by finding additional leptons from $b$
semileptonic decay (SLT tag).  In Ref.~\cite{prdl} we found a $2.8\sigma$
excess of signal over the expectation from background.
The interpretation of the excess
as top quark production was supported by a peak in the mass distribution for
fully reconstructed events.  Additional evidence was found
in the jet energy distributions in lepton+jet events\cite{kprd}.
An upper
limit on the $\ttbar$ production cross section has been published by the D0
collaboration\cite{d0}.

We report here on a data sample containing 19 $\ipb$ used in Ref.~\cite{prdl}
and 48 $\ipb$ from the current Fermilab Collider run, which began early in 1994
and is expected to continue until the end of 1995.

%

The CDF detector consists of a magnetic spectrometer surrounded by calorimeters
and muon chambers\cite{cdf_det}.  A new low-noise, radiation-hard,
four-layer silicon vertex detector, located immediately outside the beampipe,
provides
precise track reconstruction in the plane transverse to the beam and is used
to identify secondary vertices from $b$ and $c$ quark
decays\cite{svxp}.  The momenta of charged particles are measured
in the central tracking chamber (CTC), which is in a 1.4-T
superconducting solenoidal magnet. Outside the CTC, electromagnetic and
hadronic calorimeters cover the
pseudorapidity region $|\eta|<4.2$\cite{coord} and are used to identify jets
and electron candidates. The calorimeters are also used to measure
the missing transverse energy, $\met$, which can indicate the
presence of undetected energetic neutrinos. Outside the calorimeters,
drift chambers in the region $|\eta|<1.0$ provide muon identification.
A three-level trigger selects the inclusive electron and muon events
used in this analysis.  To improve the $\ttbar$ detection efficiency, triggers
based on $\MET$ are added to the lepton triggers used in Ref.~\cite{prdl}.

The data samples for both the dilepton and lepton+jets analyses are subsets of
a sample of high-$\Pt$ inclusive lepton events
that contain
an isolated electron with $\Et>20$ GeV or an isolated muon with $\Pt>20$ GeV/c
in the central region ($|\eta|<1.0$).  Events which contain a second lepton
candidate are removed as possible Z bosons if an $ee$ or $\mu\mu$ invariant
mass is between 75 and 105 GeV/c$^2$.  For the lepton+jets analysis, an
inclusive W boson sample is made by requiring \MET$>20$ GeV.  Table~\ref{njet}
classifies the W events by the number of
jets with observed $\Et>15$ GeV and $|\eta|<2.0$.  The dilepton sample
consists of inclusive lepton events that also have a second
lepton with $\Pt>20$ GeV/c, satisfying looser lepton identification
requirements.  The two leptons must have opposite electric charge.

%
The primary method for finding top quarks in the lepton+jets channel
is to search for secondary vertices from $b$ quark decay (SVX tagging).  The
vertex-finding efficiency
is significantly larger now than previously due to an improved
vertex-finding algorithm and the performance of the new
vertex detector.  The previous vertex-finding algorithm searched
for a secondary vertex with 2 or more tracks.  The new algorithm
first searches for vertices with 3 or more tracks with looser track
requirements, and if that
fails, searches for 2-track vertices using more stringent track and vertex
quality criteria.  The efficiency for tagging a $b$ quark is measured
in inclusive electron and muon samples which are enriched in $b$ decays.
The ratio of the measured efficiency to the prediction of a detailed Monte
Carlo is $0.96\pm0.07$, with good agreement ($\pm2\%$) between the electron
and muon samples.  The efficiency for tagging at
least one $b$ quark in a $\ttbar$ event with $\ge3$ jets is determined from
Monte Carlo to be $(42\pm5)\%$ in the current run, compared to the $(22\pm6)\%$
reported in the
previous publication\cite{bbug}.  In this paper we apply the new vertex finding
algorithm to the data from the previous and the current runs.

In Ref.~\cite{prdl}, we presented two methods for estimating the
background
to the top quark signal.  In method 1, the observed tag rate in
inclusive jet samples is used to calculate the background from mistags and
QCD-produced heavy quark pairs ($\bbbar$ and $\ccbar$) recoiling against a
W boson.  This is an overestimate of the background because there
are sources of heavy quarks in an inclusive jet sample that are not present in
W+jet events.  In method 2, the mistag rate is again measured with
inclusive jets, while the fraction of W+jet events that are $W\bbbar$ and
$W\ccbar$ is estimated from
a Monte Carlo sample, using measured tagging efficiencies.
In the present analysis,
we use method 2 as the best estimate of the SVX-tag background.  The improved
performance of the new vertex detector, our ability to simulate its
behavior accurately, and the agreement between the prediction and data in the
W + 1-jet and W + 2-jet samples
make this the natural choice.  The calculated background, including the small
contributions from non-W background, W$c$ production, and vector boson pair
production, is given in Table~\ref{njet}.

\begin{figure}[hp]
\epsffile[23 162 522 644]{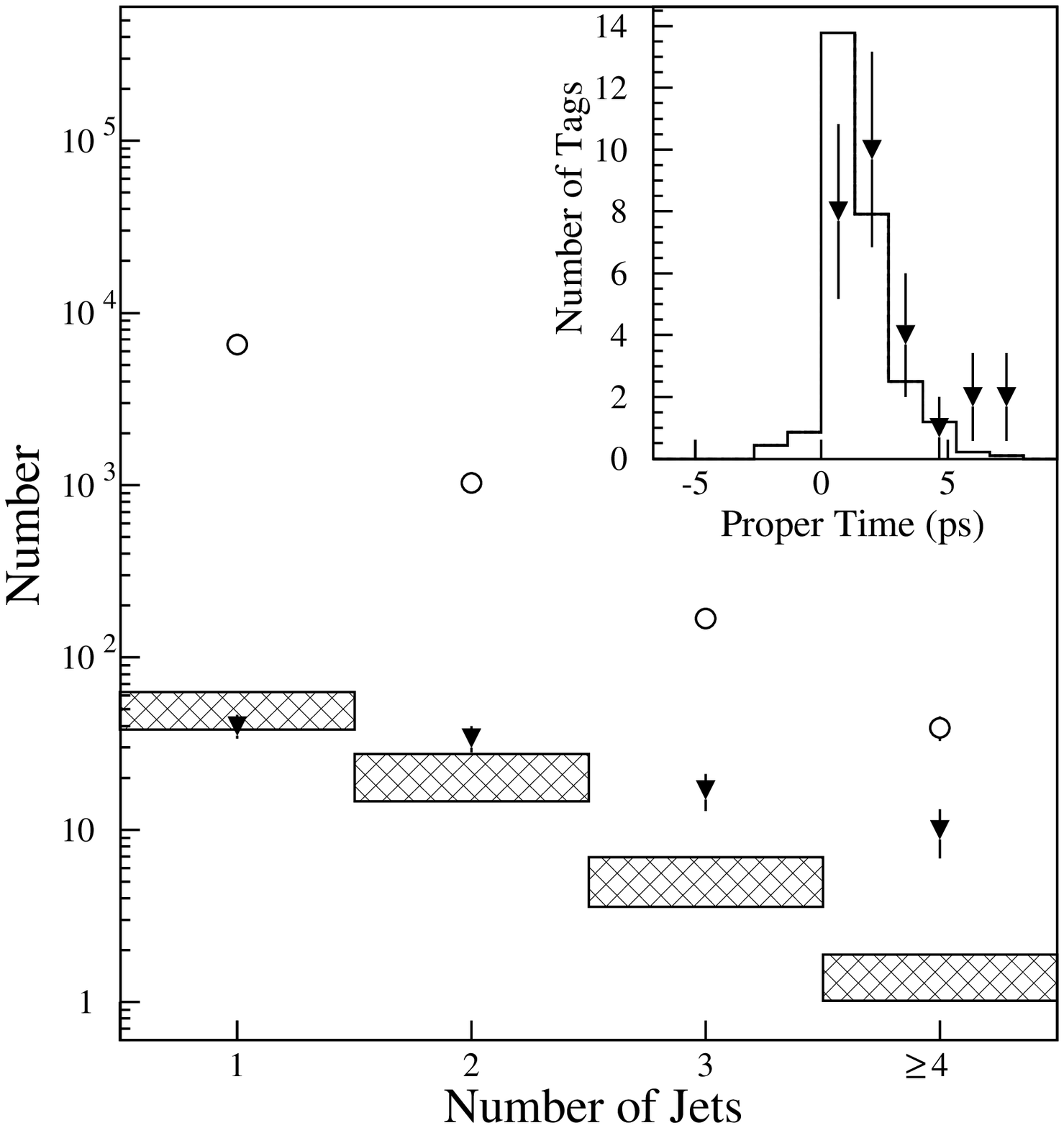}
\vspace{2.0in}
\caption{Number of events before SVX tagging (circles), number of tags
observed (triangles), and expected number of background tags (hatched) versus
jet multiplicity.  Based on the excess number of tags in events with $\ge3$
jets, we expect an additional 0.5 and 5 tags from $\ttbar$ decay in the
1 and 2 jet bins respectively.  The inset shows the secondary vertex
proper time distribution for the 27 tagged jets in the W + $\ge3$-jet
data (triangles)
compared to the expectation for $b$ quark jets from $\ttbar$ decay.}
\label{ctau}
\end{figure}

The numbers of SVX tags
in the 1-jet and 2-jet samples are consistent with the expected
background
plus a small $\ttbar$ contribution (Table~\ref{njet} and Figure~\ref{ctau}).
However for the W + $\ge3$-jet signal region, 27 tags are observed compared
to a predicted background of
$6.7\pm2.1$ tags\cite{old}.  The probability of the background
fluctuating to $\ge27$ is calculated to be $2\times10^{-5}$
(see Table~\ref{results}) using the procedure outlined in reference 1
\cite{meth1}.
The 27 tagged jets are in 21 events; the 6 events with 2 tagged jets can be
compared with 4 expected for the top+background hypothesis and $\le1$ for
background alone.
Figure~\ref{ctau} also shows the decay lifetime distribution for the SVX tags
in
W + $\ge3$-jet events.  It is consistent with the distribution predicted for
$b$ decay from the $\ttbar$ Monte Carlo simulation.  From the number of SVX
tagged events, the estimated
background, the calculated $\ttbar$ acceptance, and the integrated luminosity
of
the data sample, we calculate the $\ttbar$ production cross section to be
$6.8^{+3.6}_{-2.4}$ pb, where the uncertainty includes both statistical and
systematic effects.  This differs from the cross section given in
Ref.~\cite{prdl} by $6.9\pm5.9$ pb.

%

The second technique
for tagging $b$ quarks (SLT tagging) is to search for an additional
lepton from semileptonic $b$ decay.
Electrons and muons are found by matching
CTC tracks with electromagnetic energy clusters or tracks in the muon chambers.
To maintain acceptance for leptons coming directly from $b$ decay and from
the daughter $c$ quark, the $\Pt$ threshold is kept low (2 GeV/c).  The only
significant change to the
selection algorithm compared to Ref.~\cite{prdl} is that the
fiducial region for SLT muons has been increased from $|\eta|<0.6$ to
$|\eta|<1.0$, resulting in an increase of the SLT total
acceptance and background by a factor of 1.2.

The major backgrounds in the SLT analysis are hadrons that are misidentified as
leptons, and electrons from unidentified photon conversions.  These rates
and the smaller $W\bbbar$ and $W\ccbar$ backgrounds  are
determined directly from inclusive jet data.  The remaining
backgrounds are much smaller and are calculated using the techniques discussed
in Ref.~\cite{prdl}.  The efficiency of the algorithm is measured with photon
conversion and $J/\psi\rightarrow\mu\mu$ data.  The probability of finding an
additional $e$ or $\mu$ in a $\ttbar$ event with $\ge3$ jets is $(20\pm2)\%$.
Table~\ref{results}
shows the background and number of observed tags for the signal region
($W+\ge3$ jets).  There are 23 tags in 22 events, with $15.4\pm2.0$ tags
expected from background.  Six events contain both an SVX and SLT tag, compared
to the expected 4 for top+background and 1 for background alone.

%

The dilepton analysis
is very similar to that previously reported\cite{prdl}, with slight
modifications
to the lepton identification requirements to make them the same as those used
in the single lepton analysis.  The dilepton data sample, described above,
is reduced by additional requirements on $\MET$ and the number of jets.
In order to suppress background from Drell-Yan lepton pairs, which have little
or no true $\MET$, the $\MET$ is corrected to account for jet energy
mismeasurement\cite{prdl}.  The magnitude of the corrected $\MET$ is required
to be at least 25 GeV and, if $\MET$ is less than 50 GeV, the azimuthal angle
between the
$\MET$ vector and the nearest lepton or jet must be greater than 20\degs.
Finally, all events are required to have at least two jets with
observed $\Et>10$ GeV and $|\eta|<2.0$.

The major backgrounds are Drell-Yan lepton pairs, $Z\rightarrow\tau\tau$,
hadrons misidentified as leptons, WW,
and $\bbbar$ production.  We calculate the first three from data and the last
two with Monte Carlo simulation\cite{prdl}.  As shown in Table~\ref{results}
the total background expected is
$1.3\pm0.3$ events.  We observe a total of 7 events, 5 $e\mu$ and
2 $\mu\mu$.  The relative numbers are consistent with our dilepton acceptance,
60\% of which is in the $e\mu$ channel.  Although we have estimated the
expected background from
radiative Z decay to be small (0.04 event), one of the $\mu\mu$ events contains
an energetic photon with a $\mu\mu\gamma$ invariant mass of 86 GeV/c$^2$.
To be conservative, we have removed that event from the final sample,
which thus contains 6 events.  Three of these events contain a total of 5
$b$-tags, compared with an expected 0.5 if the events are background.  We
would expect 3.6 tags if the events are from $\ttbar$ decay.  When the
requirement that the leptons have opposite charge is
relaxed, we find one same-sign dilepton event ($e\mu$) that passes all
the other event selection criteria.  The expected number of same sign events
is 0.5, of which 0.3 is due to background and 0.2 to $\ttbar$ decay.

%

In summary, we find 37 $b$-tagged $W+\ge3$-jet events that contain 27 SVX tags
compared to $6.7\pm2.1$ expected from background and 23 SLT tags with an
estimated background of $15.4\pm2.0$.  There are 6 dilepton events compared to
$1.3\pm0.3$ events expected from background.  We have taken the product (P) of
the three probabilities in Table~\ref{results}
and calculated the likelihood that a fluctuation of the background alone would
yield a value of P no larger than that which we observe.  The result is
$1\times10^{-6}$, which is equivalent to a $4.8\sigma$ deviation in a Gaussian
distribution\cite{sig}.
Based on the excess number of SVX tagged events, we expect an excess of 7.8
SLT tags and 3.5 dilepton
events from $\ttbar$ production, in good agreement with the observed numbers.

We have performed a number of checks of this analysis.  A good control sample
for $b$-tagging is Z+jet events, where no top contribution is expected.  We
observe 15, 3, and 2 tags (SVX and SLT) in the Z + 1-jet, 2-jet, and $\ge3$-jet
samples respectively, compared with the background predictions of 17.5, 4.2,
and 1.5.  The excess over background that was seen in Ref.~\cite{prdl}
is no longer present.  In addition, there is no discrepancy between the
measured and predicted W + 4-jet background, in contrast to a small deficit
described in reference 1 \cite{as}.

%

Single lepton events with 4 or more jets can be kinematically reconstructed to
the $\ttbar\rightarrow WbW\bar{b}$ hypothesis, yielding for each event an
estimate of the top quark mass\cite{prdl}.  The lepton, neutrino ($\MET$), and
the four highest-$\Et$ jets are assumed to be the $\ttbar$ daughters\cite{cor}.
There are multiple
solutions, due to both the quadratic ambiguity in determining the longitudinal
momentum of the neutrino and the assignment of jets to the parent W's and
$b$'s. For each event, the solution with the lowest fit $\chi^2$ is
chosen.
Starting with the 203 events with $\ge3$ jets, we require each event to have
a fourth jet with $\Et>8$ GeV and $|\eta|<2.4$.  This yields a sample of 99
events, of which 88 pass a loose $\chi^2$ requirement on the fit.  The
mass distribution for these events is shown in Figure~\ref{mpretag}.  The
distribution is consistent with the predicted mix of approximately 30\%
$\ttbar$ signal and 70\%
W+jets background.  The Monte Carlo background shape agrees well with that
measured in a
limited-statistics sample of Z+4-jet events as well as in a QCD sample selected
to approximate non-W background.  After requiring an SVX or SLT $b$-tag,
19 of the events remain, of which $6.9^{+2.5}_{-1.9}$ are expected to be
background.  For these events, only solutions in which the tagged jet is
assigned to one of the $b$ quarks are considered.  Figure~\ref{mtag} shows
the mass distribution for the tagged events. The mass distribution in the
current run is very similar to that from the previous run.  Furthermore,
we have employed several mass fitting techniques which give nearly
identical results.

To find the most likely top mass, we fit the mass distribution to a sum of
the expected distributions from the W+jets background and a top quark of mass
$\Mt$\cite{prdl}.  The --ln(likelihood) distribution from the fit is shown in
the Figure~\ref{mtag} inset.  The best fit mass is 176 GeV/c$^2$ with a
$\pm8$ GeV/c$^2$ statistical
uncertainty.  We make a conservative extrapolation of the systematic
uncertainty from our previous publication, giving
$M_{top}=176\pm8\pm10$ GeV/c$^2$.  Further studies of systematic uncertainties
are in progress.

The shape of the mass peak in
Figure~\ref{mtag} provides additional evidence for top quark production,
since the number of observed $b$-tags is independent of the observed mass
distribution.  After including systematic effects in the predicted background
shape, we find a $2\times10^{-2}$ probability that the observed mass
distribution is consistent with the background (Kolmogorov-Smirnov test).
This is a conservative measure because it does not explicitly take into
account the observed narrow mass peak.

%

In conclusion, additional data confirm
the top quark evidence presented in Ref.~\cite{prdl}.
There is
now a large excess in the signal that is inconsistent with the
background prediction by $4.8\sigma$,
and a mass distribution with a $2\times10^{-2}$ probability of being consistent
with the background shape.  When combined, the signal size and mass
distribution
have a $3.7\times10^{-7}$ probability of satisfying the background hypothesis
($5.0\sigma$).
In addition, a substantial fraction of the jets in the dilepton events are
$b$-tagged.  This establishes the existence of the top quark.
The preliminary mass and cross section measurements yield $\Mt=176\pm8\pm10$
GeV/c$^2$ and $\sigma_{\ttbar}=6.8^{+3.6}_{-2.4}$ pb.

%

This work would not have been possible without the skill and
hard work of the Fermilab staff. We thank
the staffs of our institutions for their many contributions
to the construction of the detector.
This work is supported by the U.S. Department of
Energy, the National Science Foundation, the Natural Sciences and
Engineering Research Council of Canada,
the Istituto Nazionale di Fisica Nucleare of Italy, the
Ministry of Education, Science and Culture of Japan,
the National Science Council of the Republic of China,
and the A.P. Sloan Foundation.

%

\newpage
\singlespacing
\begin{samepage}

\end{samepage}

\doublespacing

\newpage
\begin{samepage}
\begin{table}[h]
\begin{center}
\vspace{0.25in}
\begin{tabular}{lrrr}
\hline\hline
\multicolumn{1}{c|}{\ } &\multicolumn{1}{c|}{observed}
   &\multicolumn{1}{c|}{observed}  &\multicolumn{1}{c}{background} \\
\multicolumn{1}{c|}{N$_{\rm jet}$}  &\multicolumn{1}{c|}{events}
&\multicolumn{1}{c|}{SVX tags} &\multicolumn{1}{c}{tags expected} \\
\hline
   1        &   6578 &   40  &  $50\pm12$  \\
   2        &   1026 &   34  &  $21.2\pm6.5$ \\
   3        &    164 &   17  &  $5.2\pm1.7$ \\
 $\geq$~4   &     39 &   10  &  $1.5\pm0.4$ \\
\hline\hline
\end{tabular}
\end{center}
\caption{Number of lepton+jet events in the 67 $\ipb$ data sample along with
the numbers of SVX tags observed and the estimated background.
Based on the excess number of tags in events with $\ge3$ jets, we expect
an additional 0.5 and 5 tags from $\ttbar$ decay in the
1 and 2 jet bins respectively.}
\label{njet}
\end{table}
\begin{table}[h]
\begin{center}
\vspace{0.25in}
\begin{tabular}{lccc}
\hline \hline
 \multicolumn{1}{c}{Channel:} & SVX & SLT &  Dilepton \\ \hline
 observed & 27 tags & 23 tags & 6 events \\
 expected background & $6.7 \pm 2.1$ & $15.4 \pm 2.0$ & $1.3 \pm 0.3$ \\
 background probability & $2\times10^{-5}$ & $6\times10^{-2}$
   & $3\times10^{-3}$ \\ \hline \hline
\end{tabular}
\end{center}
\caption{The numbers of tags or events observed in the three channels along
with the expected background and the probability that the background would
fluctuate to the observed number or more.}
\label{results}
\end{table}
\end{samepage}
%
\newpage
\newpage
\begin{figure}[h]
\gepsfcentered[23 162 522 644]{pretag_cons.PS}
\caption{Reconstructed mass distribution for the W + $\ge4$-jet sample prior to
$b$-tagging (solid).  Also shown is the background distribution (shaded),
with the normalization constrained to the calculated value.}
\label{mpretag}
\end{figure}
%
%
\newpage
\begin{figure}[h]
\gepsfcentered[23 162 522 644]{tag_cons_like.PS}
\caption{Reconstructed mass distribution for the $b$-tagged W + $\ge4$-jet
events (solid).  Also shown are the background
shape (dotted) and the sum of background plus $\ttbar$ Monte Carlo for
$M_{top} = 175$~GeV/c$^2$ (dashed), with the background constrained to the
calculated value, $6.9^{+2.5}_{-1.9}$ events.
The inset shows the likelihood fit used to
determine the top mass.}
\label{mtag}
\end{figure}
%
%

\begin{thebibliography}{99}
\bibitem{coord} In the CDF coordinate system, $\theta$ is the polar angle with
respect to the proton beam direction. The pseudorapidity, $\eta$, is defined as
$-\ln\tan\frac{\theta}{2}$.  The transverse momentum of a particle is
$P_T = P\sin\theta$.  If the magnitude of this vector is obtained using the
calorimeter energy rather than the spectrometer momentum, it becomes the
transverse energy ($E_T$).  The difference between the vector sum of all the
transverse energies in an event and zero is the missing transverse energy
(\MET).
\bibitem{old} For comparison we note that if we had used both the tagging
algorithm and background calculation (method 1)
presented in Ref.~\cite{prdl}, we would have 24 observed tags with
a predicted background of $8.8\pm0.6$ tags.
\bibitem{meth1} We get essentially the same probability
if we use method 1 for the SVX tag background because of its smaller
systematic uncertainty.
\bibitem{sig} This technique is chosen because we are combining channels with
very different expected background rates.  For comparison, if we apply the
method used in Ref.~\cite{prdl} to the SVX and dilepton channels, the two
low background modes, we obtain a probability of $1.5\times10^{-6}$.
\bibitem{as}  The improved agreement is due to the smaller $\ttbar$ production
cross section obtained in this analysis as well as correcting an overestimate
in
Ref.~\cite{prdl} in the Monte Carlo background prediction.
\bibitem{cor} The jet energies used in the mass fitting have been corrected
for instrumental and fragmentation effects.
\end{thebibliography}
\end{document}